\documentclass[submission,copyright,creativecommons]{eptcs}
\usepackage{breakurl}            
\usepackage{underscore}          
\usepackage[inline]{enumitem}
\usepackage{enumitem}
\usepackage{booktabs}
\usepackage{cite}
\usepackage{hyperref}
\usepackage{cleveref}
\usepackage{paralist} 
\usepackage{xcolor}
\setlist[description]{topsep=0pt,itemsep=-1ex,partopsep=1ex,parsep=1ex}
\makeatletter

\title{Exploring Usable Security to Improve the Impact of Formal Verification: A Research Agenda}

\author{%
\makebox[.3\linewidth]{Carolina Carreira}
\institute{INESC-ID and IST, \\ University of Lisbon, Portugal}
\and
\makebox[.3\linewidth]{João F. Ferreira}
\institute{INESC-ID and IST, \\ University of Lisbon, Portugal}
\and
\makebox[.3\linewidth]{Alexandra Mendes}
\institute{HASLab, INESC TEC and \\Universidade da Beira Interior, Portugal}
\and
\makebox[.3\linewidth]{Nicolas Christin}
\institute{Carnegie Mellon University \\Pittsburgh, Pennsylvania, USA}
}

\providecommand{\event}{AppFM 2021}
\def\titlerunning{Exploring Usable Security to Improve the Impact of Formal Verification}
\def\authorrunning{C. Carreira et al.}

\begin{document}
\maketitle

\begin{abstract}
As software becomes more complex and assumes an even greater role in our lives, formal verification is set to become the gold standard in securing software systems into the future, since it can guarantee the absence of errors and entire classes of attack. Recent advances in formal verification are being used to secure everything from unmanned drones to the internet. 

At the same time, the usable security research community has made huge progress in improving the usability of security products and end-users 
comprehension of security issues. However, there have been no human-centered studies focused on the impact of formal verification on 
the use and adoption of formally verified software products.
We propose a research agenda to fill this gap and to contribute with the first collection of studies on people’s mental models on formal verification and associated security and privacy guarantees and threats. The proposed research has the potential to increase the adoption of more secure products and it
can be directly used by the security and formal methods communities to create more effective and secure software tools.
\end{abstract}

\section{Introduction}
Formal verification can be used to secure software systems by guaranteeing the absence of errors and entire classes of attack --- recent advances are being used to secure everything from unmanned drones to the internet~\cite{fisher2017hacms}.
However, formally guaranteeing security properties does not guarantee that end-users will trust the verified systems. Since the seminal paper by Whitten and Tygar~\cite{whitten1999johnny}, usability problems --- defined to include human factors, such as mental models 
--- have been identified as a major factor in users disregarding security mechanisms. For example, despite some experts recommending password managers, the adoption of these tools is still low, partly because users distrust them~\cite{pearman2019people}.

While progress has been made in improving the usability of security products and end-users comprehension of security issues, there have been no human-centered studies focused on the impact of formal verification on the use and adoption of formally verified software products.
Given the recent and increasing industrial adoption of formal verification, the time is ripe to perform 
these studies.
The outcomes of the proposed research can increase the adoption of formal verification to secure software systems and change users’ behaviors when it comes to adopting security technologies. 

In this research agenda, we expose this gap in knowledge and propose several future research goals in three parts. First, we discuss the current perception that users have of formally verified products, secondly how to shape these perceptions, and, thirdly, research methodologies. We conclude our presentation by exposing specific examples of two important domains that have a massive user base (passwords) or are projected to grow immensely (cryptocurrency systems and DeFi protocols), and that may benefit from the proposed research.

\section{Background and Current Problems}
Security normally adds complexity to a system and interferes with the user’s primary goals. 
Two-factor authentication is a good example: it adds complexity to the authentication process and is strongly recommended for bank applications, but some users will be unsatisfied if such procedures are required for unimportant accounts~\cite{shneiderman2016designing}.
Whitten and Tygar~\cite{whitten1999johnny} analyzed email encryption as a usability problem in their seminal paper written in 1999 and helped establish the area commonly known as “usable security”. In the last decade, there has been a lot of work on usable security, across many domains~\cite{mai2020user,pearman2019people,ion2015no}.

On the other hand, formal verification has been used to guarantee  security properties of many software systems. Well-known examples include correctness and security verification of OpenSSL HMAC~\cite{beringer2015verified} and HMAC-DRBG~\cite{ye2017verified}, verification of TLS and other components of HTTPS~\cite{bhargavan2017everest}, and high-assurance software for unmanned aerial vehicles~\cite{fisher2017hacms}.

Despite all these developments, there have been no human-centered studies focused on the impact of formal verification on the use and adoption of formally verified software products. There is a
wide array of factors that influence users' adoption, retention, and usage of software. These factors can be a source of problems or undesired user behaviors. They include, among others: 
\begin{enumerate*}[label=\arabic*)]

    \item \textbf{Perceptions}, being in regards to perceived privacy, usability or satisfaction (e.g., Pearman et al.~\cite{pearman2019people} cited perceptions of increased security as a reason for the adoption of password managers);

    \item People commonly construct implicit \textbf{mental maps} to understand complex systems when the systems’ functionality goes beyond their technical knowledge~\cite{kearney1997toward}, and such tacit knowledge influences human behavior, even unknowingly; 

    \item A products' \textbf{usability} can impact \textbf{adoption} rates (e.g. a reason for low adoption of security software in the past has been usability problems~\cite{pearman2019people});

    \item Lack of \textbf{knowledge} about a product can provide a false sense of security or
    discourage users from using it: for example, Gao et al.~\cite{gao2015two} found that bitcoin users' lack of knowledge about the technology presented an entry barrier for new users; 
    
    \item  \textbf{Motivation} is also a crucial factor in encouraging users to overcome the initial overhead of using new software~\cite{shneiderman2016designing}.  
\end{enumerate*}
 
It is known that using a well-designed security mechanism is still more effort than not using it at all, and users will always be tempted to cut corners~\cite{sasse2005usable}. Hence, it is the developers' and designers' job to convince users to use their products.
We thus believe that human-centered studies regarding end-users mental models, cognitive biases, common misconceptions, and general perceptions of formal verification are necessary to improve the adoption and impact of formally verified products.

\section{Research Agenda}
In this section, we discuss three main future research paths for human-centered studies on formal verification. First, we propose studying the current perception users have of formally verified products, secondly how to shape this perception, and thirdly relevant research methodologies. For each path, we present the main research questions to be addressed.

\subsection{Understanding the impact of formal verification}
As users implicitly construct mental models of systems~\cite{kearney1997toward}, their behavior is affected,
even unknowingly. We claim that performing a thorough study of mental models on formal verification, similar to studies on mental models about other technologies~\cite{mai2020user}, is necessary and important to understand users' misconceptions and their grasp of what it means for a product to have formally verified features.

Formal verification of software can guarantee the absence of errors and entire classes of attack. This is a major selling point of formally verified software and can affect users' choices when deciding what software to use. As such, it is important to understand users' perceptions of formal verification and whether it has an impact on their adoption of software products.
Additionally, users may value the use of formal verification in some types of products, but not in others.
For example, in a previous study, secure access to financial accounts was valued above other types of online accounts~\cite{pearman2019people}. It is thus important to survey in what type of software products formal verification has a greater impact on users.
 
Misconceptions about underlying concepts and processes are common and can have serious effects on trust and adoption of products. For example, Pearman et al.~\cite{pearman2019people} studied users' perceptions of password managers and found that lack of knowledge about how they worked posed a barrier to their effective use and adoption. In a different study, Mai et al.~\cite{mai2020user} studied users' mental models of cryptocurrency systems and found a misconception where many participants presumed their transactions could not be tracked due to the encryption used; this is not true and gave users a false sense of privacy. 
If users hold misconceptions about the role of formal verification in a system, these need to be identified and addressed. 
It is also relevant to know if users find that formal verification makes products safer.

Learning more about users' perceptions of formal verification is crucial for formal methods practitioners, as with this information they can better understand the impact their products and developments have on the end-user. With this knowledge, it is easier to convey the \textbf{usefulness} of formal verification.

The main research questions to be addressed in this line of work are:

\medskip

\begin{center}
\noindent\fcolorbox{black}{black!10}{
\parbox{0.95\textwidth}{%
    \textbf{Main research questions:}
    \begin{compactitem}
          \item How do users perceive formal verification in software products?
          \item What are users' misconceptions about formally verified products?
          \item What is the impact of formal verification on the adoption of software products?
          \item In what type of software products does formal verification have a larger impact on users?
    \end{compactitem}
}%
}
\end{center}

\subsection{Shaping users' perception of formal verification}
After learning about users' current perceptions of formal verification concepts, 
we can address identified issues and misconceptions.
As stated before, usability problems have been proved to be a barrier to user adoption of software products and users have a hard time understanding security concepts or how to be secure. 
For example, despite experts recommending password managers, some users report unawareness of their existence as a reason to not use them, and even when users are aware of their existence, some are still reluctant in using them due to a lack of understanding of their security properties~\cite{pearman2019people}. 
This lack of knowledge can have a significant impact: as mentioned above, Mai et al.~\cite{mai2020user} found that a poor understanding of the cryptocurrency system being used exposes users to privacy risks. 

We claim that it is important to \textbf{convey information} to users to make them aware of the existence of formal verification and the advantages it can provide. Knowing its advantages can motivate users to choose formally verified products.
This information can be general (e.g. educating users about what formal verification is) or domain-specific (e.g. helping users understand what specific features are formally verified and what it means for them).

From the perspective of the user interface, several approaches can be taken to convey important information. Examples include the use of support/help tools like tooltips, tutorials, informative pop-ups, icons, wiki pages, and frequently asked questions pages~\cite{shneiderman2016designing, tidwell2010designing}. For example, in the PassCert project\footnote{The PassCert project aims to build a formally verified PM and to investigate ways to effectively convey to users the formally verified properties. Project URL: \url{https://passcert-project.github.io}}, we are exploring all of the above to convey to users what properties are guaranteed and what value it brings to them.
An important point is to be careful to not overstate the formal guarantees and to ensure that users understand that a formally verified product might still be vulnerable.
From a more general perspective, there is an opportunity to 
raise awareness through advocacy in schools and the media. 
Educating users, by explaining formal verification usage and associated security and privacy guarantees, or by describing the dangers of unsafe software, could serve as motivation for the adoption and long-term retention of users in formally verified software. The study of what are the best ways to effectively convey formal verification concepts to users and how this information impacts their adoption is crucial and valuable to the formal methods community.

Regarding implicit knowledge maps, the construction of adequate mental models about formal verification should be enabled by the 
suggestions described above,
but there are other generic usability techniques that should not be disregarded, such as
the implementation of clear navigation systems and other interface design choices (e.g. following the Eight Golden Rules by Shneiderman et al.~\cite{shneiderman2016designing}). A survey of specific usability principles that should be applied to formally verified products is missing. 
Usability can be used to enable users' understanding of formal methods topics and ensure the end-product is usable, this allied with a well-implemented interface can influence users' perceptions on formal verification.

The main research questions to be addressed in this line of work are:
\medskip

\begin{center}
\noindent\fcolorbox{black}{black!10}{
\parbox{0.95\textwidth}{%
    \textbf{Main research questions:}
    \begin{compactitem}
          \item How can we effectively convey formal verification concepts to users?
          \item Does users' understanding of formal verification concepts have an impact on the adoption of formally verified software?
          \item What usability principles should be followed to ensure adequate mental models of formal verification?
    \end{compactitem}
}%
}
\end{center}

\subsection{User testing methodologies}

We now turn our attention to \textbf{how} we can acquire information on users' mental models, perceptions, and understanding of formal verification. For this, existing techniques can be used, such as:
\begin{enumerate*}[label=\arabic*)]
    \item \textbf{Questionnaires}: 
    several studies regarding users' perceptions of software products have used this method with success~\cite{pearman2019people, presthus2017motivations, ion2015no}. Questionnaires can also be combined with other user testing tools such as interviews and usability tests~\cite{shneiderman2016designing, tidwell2010designing}. When scaled up to encompass a large number of users using online services, questionnaires are a strong quantitative research tool~\cite{tidwell2010designing}.
    \item \textbf{Interviews}: these can provide detailed and qualitative information. They can be structured or non-structured and are also commonly used~\cite{pearman2019people, ion2015no, gao2015two}. One-on-one interviews offer more qualitative results that can serve as the basis for understanding users' expectations, vocabulary, goals, and perceptions~\cite{tidwell2010designing}.
\end{enumerate*} 
Both techniques can and should be adapted to particular products or testing goals and their outputs could be used both to better understand users' perceptions of formal verification and as feedback about a possible formally verified product. A suggestion we give is to survey these methodologies to understand how to better apply them and understand the impact of formally verified software. 
Biases that might affect results need to be considered, such as
the Hawthorne effect where users may be inclined to agree with the researchers~\cite{merrett2006reflections}. 

As stated in the previous section, it is also relevant to learn about the usability of any formally verified product. An iterative design process, with several rounds of user testing, is recommended~\cite{shneiderman2016designing}. Within this process several methods can be used, for example:
\begin{compactitem}
      \item \textit{By experts} -- heuristic evaluations, cognitive walkthroughs where experts compare the interface with a set of heuristics rules and simulate users walking through the interface; 
      \item With users \textit{during development} -- early user studies can be done with prototypes, by asking users to perform tasks while \textit{thinking aloud} and iterate over that feedback; other tests include competitive testing to compare different interface versions. Nowadays, remote usability testing has gained popularity as it can be done to a large number of users through online communities including Amazon Mechanical Turk~\cite{shneiderman2016designing}.
      \item With users during \textit{active use} -- interviews and focus-group discussions can be productive because the interviewer can pursue specific issues of concern to help in better understanding the users' perspectives~\cite{shneiderman2016designing}. Software should also provide developers with a continuous data logging of user performance and supply information about patterns of use.
\end{compactitem}

These are a few examples of evaluation methods and techniques that can be used. Typical usability tests measure how well the user is able to perform certain tasks in the interface, and its ease of use. Here, we need user studies that go further and provide information about users' perception, understanding, and retention when considering formal methods aspects.
The \textbf{best methods to learn users' perception of formal verification and how it can affect usability are still unknown} and we claim that 
there is a need for building solid and replicable user research methodologies that meet our specific goals, provide useful data, and 
that can be applied to different domains where formal verification is used.

The main research questions to be addressed in this line of work are:

\medskip

\begin{center}
\noindent\fcolorbox{black}{black!10}{
\parbox{0.95\textwidth}{%
    \textbf{Main research questions:}
    \begin{compactitem}
          \item How can we effectively test users' understanding of formal verification?
          \item What tools can be used to learn about the usability of formally verified products?
    \end{compactitem}
}%
}
\end{center}

\section{Promising Next Steps}
Next steps for this research include the application of the previously described generic research questions to concrete domains. 
We believe that any domain where formal methods are currently being applied can be considered.
However, to maximize impact, we propose two important specific domains: one that has a massive user base (password security) and another that is projected to grow immensely in the near future (cryptocurrency systems and DeFi protocols). In this section, we describe these two domains and, for each, we provide background research, motivation, and research suggestions. 
It is important to note that these are just two examples from a large set of possible future research domains.

\subsection{Password security}
Text passwords are one of the most used security mechanisms. However, users often struggle to remember passwords~\cite{inglesant2010true,shay2014can, yan2004password}. This results in users frequently reusing passwords across multiple accounts~\cite{lyastani2018better}. Also, entering long or complex passwords on mobile devices leads users to use weak passwords~\cite{ur2015added}.
Experts recommend password managers with random-generation features to help users employ strong and unique passwords. 
However, studies have shown that password managers have low adoption rates, especially among non-experts~\cite{ion2015no}. Many users do not trust password managers~\cite{silver2014password} and several usability problems have been found~\cite{pearman2019people}.

Following previous work on formal methods applied to password security~\cite{ferreira2017certified,johnson2020skeptic}, a verified password manager that assures properties on data storage and password generation is being built in the context of PassCert ~\cite{grilo2021towards}, 
where we are currently studying the effectiveness
of status symbols and tooltips that explicitly indicate that certain actions or properties are formally assured. One of the end goals is to propose ideas to integrate non-obtrusive information and documentation on formal verification in password managers. We propose an expansion of this work by:
\begin{enumerate*}[label=\arabic*)]
    \item studying in detail users' mental models and common misconceptions on both formal verification and password managers;
    \item studying the impact of formal verification on user acquisition/adoption in this domain;
    \item applying relevant usability tools (e.g. status icons) to improve communication on the assurances provided by formal verification;
    \item performing large-scale studies that can provide statistical significant actionable findings that can increase the adoption of important tools such as password managers; and
    \item surveying user research methodologies to achieve the previously mentioned qualitative and quantitative research.
\end{enumerate*}

\subsection{Decentralized finance}
Of the many recent developments on formal verification of smart contracts~\cite{tolmach2021survey}, we highlight KEVM, a K semantics of the Ethereum Virtual Machine~\cite{hildenbrandt2018kevm}. Runtime Verification Inc is currently using K to verify the Maker Multi-Collateral Dai Protocol\footnote{https://security.makerdao.com/audit-reports\#runtime-verification-specification}. Experts noted that formal verification eliminated many of the ‘low-hanging fruit’ vulnerabilities, and recommended continued use. This suggests that DeFi companies will increasingly use formal verification, providing extra motivation for the research we propose.
A DeFi system enables financial services through decentralized peer-to-peer networks. Advantages include decentralization and transparency~\cite{chen2020blockchain}. DeFi protocols have seen staggering growth recently, with 2020 hitting the milestone figure of over \$12 billion in total locked assets.

Poor usability and lack of knowledge are major contributors to cryptocurrency security failures~\cite{krombholz2016other}. Mai et al.~\cite{mai2020user} studied users' mental models of cryptocurrency systems and found flaws and inconsistencies that expose users to security and privacy risks.
Gao et al.~\cite{gao2015two} examined adopters and non-adopters’ perceptions of bitcoin, and found that both lacked knowledge about it. Non-adopters also questioned the security of these systems~\cite{presthus2017motivations}.

Regarding this domain, as a next step, we propose to consider the Maker Protocol and Runtime Verification Inc's formalization.
Since the Maker Protocol is complex, there are two main aspects on which we propose to focus: first, investigation of usability principles that should be followed to ensure that financial tools based on the Maker Protocol can help build adequate mental models; second, survey effective ways to convey that the protocol is formally verified and measure if (and how) this can lead to wider adoption of cryptocurrency-based products. 
We propose an exploration and assessment of different alternatives to inform the users about what properties are formally verified (e.g. by adding cues and visualizations). Moreover, we propose to integrate
specialized information viewers that show (and explain) the classes of problems that were proved impossible to occur, thus helping users to make informed decisions.

\section{Conclusion}
We propose a research agenda that can create new knowledge on how users perceive critical security issues and how they perceive the value of formal verification in software security. Moreover, this knowledge can enable effective communication on the assurances provided by formal verification. As a result, its outcomes can be directly used by the security and formal methods communities to create more effective and secure software tools in most areas where formal methods are applied. 

\paragraph{\textbf{Acknowledgments.}}
We thank the anonymous reviewers for their valuable and constructive comments.
This work was partially funded by the PassCert project, a CMU Portugal Exploratory Project funded by Funda\c{c}\~ao para a Ci\^encia e Tecnologia (FCT), with reference CMU/TIC/0006/2019 and supported by national funds through FCT under project UIDB/50021/2020. 

\bibliographystyle{eptcs}
\bibliography{main}
\end{document}